\documentclass[a4paper]{article}

\usepackage{INTERSPEECH2021}
\usepackage{tabularx}
\newcommand\blfootnote[1]{%
  \begingroup
  \renewcommand\thefootnote{}\footnote{#1}%
  \addtocounter{footnote}{-1}%
  \endgroup
}

\title{Speech Denoising with Auditory Models}
\name{Mark R. Saddler$^{1,2,3,\dagger}$, Andrew Francl$^{1,2,3,\dagger}$, Jenelle Feather$^{1,2,3}$,\\ Kaizhi Qian$^{4}$, Yang Zhang$^{4}$, Josh H. McDermott$^{1,2,3}$}
\address{
  $^1$ Department of Brain and Cognitive Sciences, MIT, USA\\
  $^2$ McGovern Institute for Brain Research, MIT, USA\\
  $^3$ Center for Brains, Minds and Machines, MIT, USA\\
  $^4$ MIT-IBM Watson AI Lab, IBM Research, USA}
\email{\{msaddler, francl, jhm\}@mit.edu}

\begin{document}

\maketitle

\begin{abstract}
Contemporary speech enhancement predominantly relies on audio transforms that are trained to reconstruct a clean speech waveform. The development of high-performing neural network sound recognition systems has raised the possibility of using deep feature representations as ‘perceptual’ losses with which to train denoising systems. We explored their utility by first training deep neural networks to classify either spoken words or environmental sounds from audio. We then trained an audio transform to map noisy speech to an audio waveform that minimized the difference in the deep feature representations between the output audio and the corresponding clean audio. The resulting transforms removed noise substantially better than baseline methods trained to reconstruct clean waveforms, and also outperformed previous methods using deep feature losses. However, a similar benefit was obtained simply by using losses derived from the filter bank inputs to the deep networks. The results show that deep features can guide speech enhancement, but suggest that they do not yet outperform simple alternatives that do not involve learned features.
\end{abstract}

\section{Introduction}
\label{sec:intro}

Recent advances in speech enhancement have been driven by neural network models trained to reconstruct speech sample-by-sample \cite{park2016fully,pascual2017segan,Wang2017review,qian2017speech,rethage2018wavenet,luo2019conv,pandey2019new,Hsieh2020}. These methods provide substantial benefits over previous approaches, but nonetheless leave room for improvement. The resulting processed speech usually contains audible artifacts, and noise removal is usually incomplete at lower SNRs.

A parallel line of work has explored the use of deep artificial neural networks as models of sensory systems \cite{yamins2016using,kell2019}. Although substantial discrepancies remain \cite{rajalingham2018large, Feather2019}, such trained neural networks currently provide the best predictive models of brain responses and behavior in both the visual and auditory systems \cite{yamins2016using,kell2018task}. The apparent similarities between deep supervised feature representations and representations in the brain raises the possibility that such representations could be used as perceptual metrics. Such metrics have been successfully employed in image processing \cite{Zhang2018}, but are not widely used in audio applications.

Deep feature losses for denoising were previously proposed in \cite{Germain2019,Pilarczyk2019,su2020hifi,hsieh2020improving,kataria2020perceptual}, but were explored only for relatively high signal-to-noise ratios (SNRs), a single task and network, or were not compared to baseline methods using the same transform architecture. Additionally, direct comparisons have not been made to simpler losses derived from conventional filter banks. It was thus unclear the extent to which deep feature losses could improve on simpler approaches, and what choices in the feature training would produce the best results. The goal of this paper was to directly compare deep perceptual losses to alternative losses, and to explore the conditions in which benefits might be achieved. We found that deep feature losses produced more natural denoising compared to waveform losses, but that a similar benefit could be achieved using a loss derived from standard filter bank representations.


\section{Methods}
\label{sec:methods}

\begin{figure}[t]
  \centering
  \includegraphics[width=\linewidth]{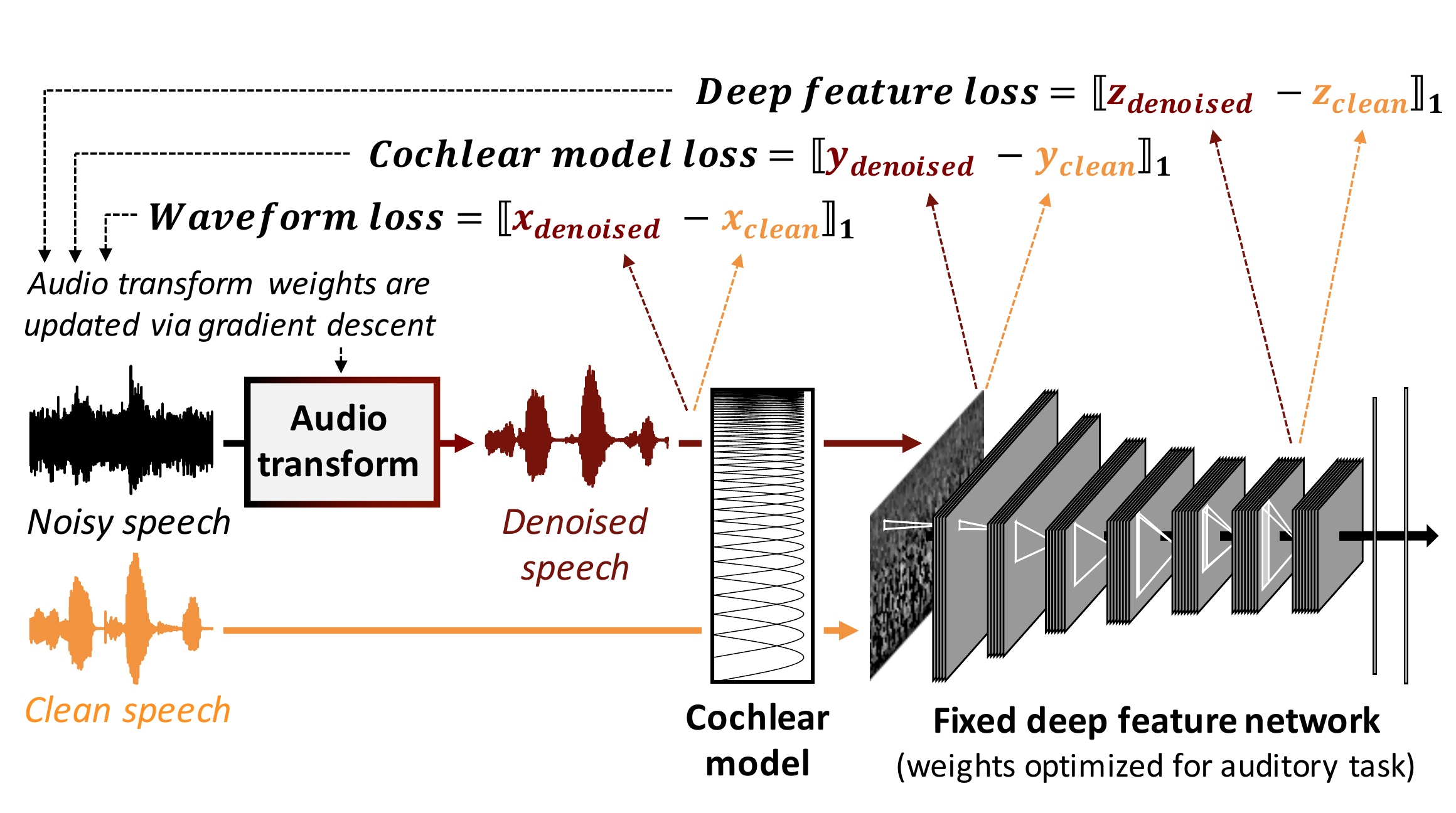}
  \caption{Schematic of audio transform training.}
  \label{fig:schematic}
\end{figure}

There were two components to our denoising approach (Figure \ref{fig:schematic}). The first component was a recognition network trained to recognize either speech or environmental sounds. Once trained, this network was used to define deep feature losses. Speech recognition is a natural choice in this context, but it also seemed plausible that more general-purpose audio features learned for environmental sound recognition might help to achieve natural-sounding audio even in speech applications. The input to the network was the output of a filter bank modeled on the human cochlea.

The second component was a waveform-to-waveform audio transform whose parameters were adjusted via gradient descent to minimize a loss function (evaluated on features of the recognition network, or the outputs of a filter bank, or on the waveform). We used a Wave-U-Net \cite{stoller2018wave}, which has been found to perform comparably to WaveNet \cite{macartney2018improved} based on objective metrics of noise reduction, but which can be specified with many fewer parameters and run with a much lower memory footprint. Code, models, and audio examples are available at: http://mcdermottlab.mit.edu/denoising/demo.html.

\subsection{Recognition Networks}
The recognition networks took as input simulated cochlear representations of 2s sound clips (audio sampled at 20 kHz). The cochlear model consisted of a bank of 40 bandpass filters whose frequency tuning mimics that of the human ear (evenly spaced on an Equivalent Rectangular Bandwidth scale \cite{glasberg1990}), followed by half-wave rectification, downsampling to 10 kHz, and 0.3 power compression \cite{mcdsim2011}.

\subsubsection{Recognition Network Architectures}
We used three feed-forward CNN architectures for the recognition networks. Each consisted of stages of convolution, rectification, batch normalization, and weighted average pooling with a hanning kernel to minimize aliasing \cite{henaff2015geodesics, Feather2019}. The three architectures were selected based on word recognition task performance from 3097 randomly-generated architectures varying in number of convolutional layers (from 4 to 8), size and shape of convolutional kernels, and extent of pooling. The selected architectures had 6 (arch1) or 7 (arch2,3) convolutional layers.

\subsubsection{Recognition Network Training}
The recognition networks were trained to perform either word recognition or environmental sound recognition. For the speech task, each training example was a speech excerpt (from the Wall Street Journal \cite{wallstreetjournal1992} or Spoken Wikipedia Corpora \cite{spokenwikipedia2016}). The task was to recognize the word overlapping with the center of the clip \cite{kell2018task,Feather2019} (out of 793 word classes sourced from 432 unique speakers, with 230,357 unique clips in the training set and 40,651 segments in the validation set). For the environmental sound recognition task, each training example was a non-speech YouTube soundtrack excerpt (from a subset of 718,625 AudioSet examples \cite{audioset2017}), and the task was to predict the AudioSet labels (spanning 516 categories in our dataset).

The three network architectures were trained on each task until performance on the validation set task plateaued. Word task classification accuracies for the three architectures were: arch1 $=90.4\%$, arch2 $=88.5\%$, and arch3 $=80.6\%$. AudioSet task AUC values were: arch1 $=0.845$, arch2 $=0.861$, and arch3 $=0.869$.

\subsection{Audio Transforms}

\subsubsection{Wave-U-Net Architecture}
The Wave-U-Net architecture was the same as in \cite{macartney2018improved}: 12 layers in the contracting path, a 1-layer bottleneck, and 12 layers in the expanding path. All layers utilized 1D convolutions with learned filters and LeakyReLU activation functions. There were 24 filters in the first layer, and the number of filters increased by a factor of 2 with each successive layer prior to the bottleneck.

\subsubsection{Deep Feature Losses}
The recognition networks were used to define a deep feature loss function as the $L1$ distance between network representations of noisy speech and clean speech. The total loss for a single recognition network and single training example was the sum of the $L1$ distances between the noisy speech and clean speech activations for each convolutional layer, weighted to approximately balance the contribution of each layer. 

\subsubsection{Cochlear Model Losses}
We also trained transforms using losses derived from the cochlear model that provided input to the recognition network, as well as variants of the model that varied in i) the number of filters (5, 10, 20, 40, 80 and 160 filters, evenly spaced on an ERB-scale \cite{glasberg1990}, with bandwidths scaled to tile the spectrum in all cases), ii) the dependence of filter bandwidth on frequency (linearly-spaced and `reversed', with broad low-frequency filters and narrow high-frequency filters, opposite to what is found in the ear), and iii) in their phase invariance (subband envelopes computed by lowpass-filtering the rectified subbands; cutoff of 100 Hz). 

\subsubsection{Wave-U-Net Training}
Out of concern that the audio transform might overfit to idiosyncrasies of any individual recognition network, we trained some transforms on losses computed simultaneously from an ensemble of three different networks (arch1,2,3), and some on just a single network (arch1).

In all cases the Wave-U-Net was trained on speech superimposed on non-speech AudioSet excerpts (the same corpora used to train the recognition networks) with SNR drawn uniformly from $[-20, +10]$ dB. AudioSet excerpts were used as the training `noise' as they were highly varied and diverse. All Wave-U-Net models were trained with the ADAM optimizer for 600,000 steps (batch size=$8$, learning rate=$10^{-4}$).

\subsubsection{Baselines}
We used two baseline models, both trained to explicitly reconstruct clean speech waveforms from noisy speech waveforms drawn from the same training set described above. The first was a previously described WaveNet \cite{rethage2018wavenet} and the second was the Wave-U-Net \cite{macartney2018improved} used with the deep feature and filter losses.

We also compared our results to those of a previously published denoising transform trained with a deep feature loss \cite{Germain2019}, using both the pre-trained model made available by Germain et al. and a Wave-U-Net that we trained on our dataset using the feature loss from \cite{Germain2019} (deep network features trained on the DCASE 2016 \cite{mesaros2016tut} environmental sound challenge).

\subsection{Evaluation}
We evaluated the trained models on 40 speech excerpts (from a separate validation set) superimposed separately on each of four types of noise signals: speech-shaped Gaussian noise, auditory scenes from the DCASE 2013 dataset \cite{giannoulis2013detection}, instrumental music from the Million Song Dataset \cite{bertin2011million}, and 8-speaker babble made from public-domain audiobooks (librivox.org). These noise sources were chosen to be distinct from those in the training set, and to span a variety of noise types to assess the generality of the trained transforms.

\subsubsection{Human Perceptual Evaluation and Objective Metrics}
We evaluated the audio transforms by conducting perceptual experiments on Amazon Mechanical Turk. Participants first completed a screening task to help ensure that they were wearing headphones or earphones \cite{Woods2017}. The participants who passed this screening task then rated the naturalness of a set of processed speech signals, presented seven at a time in a MUSHRA-like paradigm. Listeners could listen to each clip as many times as they wished and then gave each a numerical rating on a scale of 1-7. Listeners were provided with anchors corresponding to the ends of the rating scale (1 and 7). The anchor at the high end was always the original clean speech. The low-end anchor was 4-bit-quantized speech (an example of very high distortion). To help ensure that participants were using the scale as instructed, each experiment included 3 catch trials where two of the stimuli were the two anchors. In order to be included in the analysis, participants had to rate all instances of the high and low anchors as 7 and 1, respectively. 

We ran two identically structured experiments to evaluate all of our audio transforms. Experiment 1 compared various deep feature losses to baselines and contained all of the conditions listed in Table \ref{tab:expt1}. Experiment 2 compared losses derived from different cochlear filter banks and contained all of the conditions listed in Table \ref{tab:expt2}. 54 and 105 participants met the inclusion criteria for Experiments 1 and 2, respectively.

We also used three standard objective measures for evaluation: perceptual evaluation of speech quality (PESQ) \cite{recommendation2001perceptual}, short-time objective intelligibility measure (STOI) \cite{stoi}, and the signal-to-distortion ratio (SDR) \cite{vincent2006performance}.

\begin{table}[t]
  \caption{Experiment 1 results. Reported metrics are averaged across the five tested SNR levels. Higher is better for all metrics.}
  \label{tab:expt1}
  \centering
  \setlength{\tabcolsep}{2.5pt}
  \setlength{\extrarowheight}{2.5pt}
  \footnotesize
  \begin{tabularx}{\columnwidth}{
    >{\hsize=0.9\hsize\linewidth=\hsize\centering\arraybackslash}m{2.0cm}|
    >{\hsize=1.1\hsize\linewidth=\hsize\centering\arraybackslash}m{2.0cm}
    |c|ccc}
    \toprule
    \textbf{Model name} & \textbf{Loss function} & \textbf{Natural.} & \textbf{PESQ} & \textbf{STOI} & \textbf{SDR} \\
    \midrule
Cochlear model (N=40; human) & 40 ERB-spaced subbands & \textbf{4.43} & 1.55 & 0.75 & 7.16 \\
\hline
A123 & AudioSet features\newline (arch123) & \textbf{4.43} & 1.66 & 0.77 & 4.06 \\
A1+W1 & AudioSet + Word\newline features (arch1) & 4.36 & \textbf{1.68} & 0.79 & 6.18 \\
A123+W123 & AudioSet + Word\newline features (arch123) & 4.33 & 1.67 & 0.77 & 4.18 \\
A1 & AudioSet features\newline (arch1) & 4.33 & 1.65 & 0.78 & 3.63 \\
W123 & Word features\newline (archs123) & 4.24 & 1.67 & \textbf{0.79} & 6.64 \\
W1 & Word features\newline (arch1) & 4.22 & 1.63 & 0.77 & 3.30 \\
Random1 & Random features\newline (arch1) & 3.91 & 1.57 & 0.78 & 5.64 \\
Random123 & Random features\newline (arch123) & 3.84 & 1.57 & 0.77 & 5.08 \\
\hline
Germain\newline DeepFeatures & DCASE features from \cite{Germain2019} & 3.83 & 1.47 & 0.77 & 6.72 \\
Germain\newline (pre-trained) & DCASE features from \cite{Germain2019} & 2.36 & 1.14 & 0.64 & 0.93 \\
\hline
Waveform (Wave-U-Net) & Waveform & 4.17 & 1.51 & 0.76 & \textbf{7.35} \\
Waveform (WaveNet) & Waveform & 3.72 & 1.40 & 0.75 & 6.00 \\
\hline
Unprocessed input &  & 2.67 & 1.15 & 0.70 & 0.21 \\
    \bottomrule
  \end{tabularx}
\end{table}

\begin{figure}[t]
  \centering
  \includegraphics[width=\linewidth]{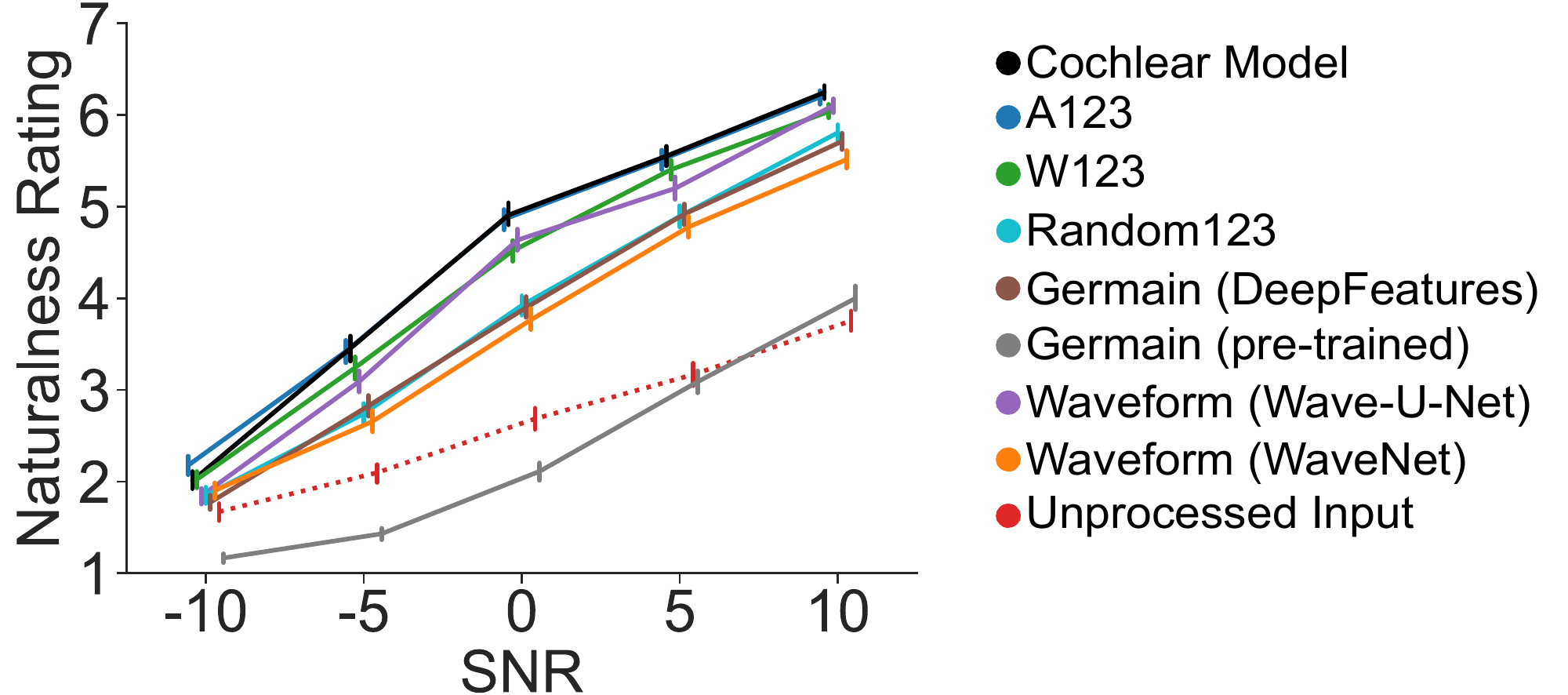}
  \caption{Rated naturalness vs. SNR for speech processed by Wave-U-Nets trained on deep feature losses, in addition to baseline models trained to reconstruct clean speech waveforms, and two versions of a related prior method \cite{Germain2019}. Error bars plot SEM (across 54 participants).}
  \label{fig:naturalness}
\end{figure}

\section{Results}
\label{sec:results}

\subsection{Deep Feature Losses Yield Improved Denoising}
The best-performing systems trained with deep perceptual feature losses outperformed both waveform-based baselines. The average objective and subjective evaluation results are shown in Table \ref{tab:expt1}. Human listeners found the speech processed by the deep feature models to be more natural than the speech processed by the baseline models. We plot the naturalness results in more detail (Figure \ref{fig:naturalness}) for two of the best-performing models trained on each of AudioSet features (A123) and word recognition features (W123), as well as a model trained on random features (Random123), the two baselines, and the two versions of the denoising network from \cite{Germain2019}.


\subsection{Learned vs. Random Deep Features}
The benefit of deep feature losses was specific to models trained with learned features. Audio transforms trained to reconstruct random features did not produce better naturalism than the baseline WaveNet, and performed worse overall than the baseline Wave-U-Net (Figure \ref{fig:naturalness}; Table \ref{tab:expt1}).

\subsection{Comparison to Previous Deep Feature Systems}
Our best-performing deep feature-based systems also outperformed previously published systems with deep feature losses. The pre-trained system from Germain et al. \cite{Germain2019} generalized poorly to our test set. Furthermore, the Wave-U-Net we trained using the deep feature loss from \cite{Germain2019} also performed worse than the baseline Wave-U-Net. These findings suggest that the features used for the perceptual loss are important, and that the DCASE task used in \cite{Germain2019} may not have produced sufficiently general features.

\subsection{Effect of Task Used to Train Deep Features}
The best results occurred for features trained on the environmental sound recognition task -- naturalism was consistently higher than for features trained on word recognition (Figure \ref{fig:naturalness}; Table \ref{tab:expt1}). However, all of the models trained with feature losses from our recognition networks produced more natural-sounding speech than the baselines, and than the systems trained with DCASE features based on \cite{Germain2019}. There was no obvious benefit from training on features from three different networks.


\begin{table}[t]
  \caption{Experiment 2 results. Reported metrics are averaged across the five tested SNR levels. Higher is better for all metrics.}
  \label{tab:expt2}
  \centering
  \setlength{\tabcolsep}{2.5pt}
  \setlength{\extrarowheight}{2.5pt}
  \footnotesize
  \begin{tabularx}{\columnwidth}{
    >{\hsize=0.9\hsize\linewidth=\hsize\centering\arraybackslash}m{2.0cm}|
    >{\hsize=1.1\hsize\linewidth=\hsize\centering\arraybackslash}m{2.0cm}
    |c|ccc}
    \toprule
    \textbf{Model name} & \textbf{Loss function} & \textbf{Natural.} & \textbf{PESQ} & \textbf{STOI} & \textbf{SDR} \\
    \midrule
Cochlear model (N=20) & 20 ERB-spaced subbands & \textbf{4.33} & 1.54 & 0.77 & \textbf{7.61} \\
Cochlear model (N=40; human) & 40 ERB-spaced subbands & 4.30 & 1.55 & 0.75 & 7.16 \\
Cochlear model (N=160) & 160 ERB-spaced subbands & 4.26 & 1.60 & 0.77 & 7.51 \\
Cochlear model (N=10) & 10 ERB-spaced subbands & 4.22 & 1.49 & 0.76 & 7.08 \\
Cochlear model (N=80) & 80 ERB-spaced subbands & 4.21 & 1.53 & 0.74 & 6.69 \\
Cochlear model (N=5) & 5 ERB-spaced subbands & 3.93 & 1.42 & 0.75 & 6.02 \\
\hline
Cochlear model (N=40; linear) & 40 linearly-\newline spaced subbands & 4.32 & 1.51 & 0.76 & 6.82 \\
Cochlear model (N=40; env.) & Envelopes of 40 ERB subbands & 4.16 & 1.59 & 0.75 & 6.94 \\
Cochlear model (N=40; reverse) & 40 reverse-ERB-\newline spaced subbands & 4.08 & 1.47 & 0.73 & 4.73 \\
\hline
A123 & AudioSet features (arch123) & 4.27 & \textbf{1.66} & \textbf{0.77} & 4.06 \\
Waveform (Wave-U-Net) & Waveform & 4.17 & 1.51 & 0.76 & 7.35 \\
\hline
Unprocessed input &  & 2.47 & 1.15 & 0.70 & 0.21 \\
    \bottomrule
  \end{tabularx}
\end{table}

\subsection{Cochlear Model Losses Match Deep Feature Losses}
Although deep features produced better performance than baselines trained using waveform losses, we found that we could reproduce their benefit using losses derived from the cochlear model inputs to the recognition networks. Based on rated naturalness, the transform trained with this `cochlear' loss performed just as well as our best model trained with deep feature losses (Table \ref{tab:expt1}). 

\subsection{Effect of Filter Bank Characteristics}
\begin{figure}[t]
  \centering
  \includegraphics[width=\linewidth]{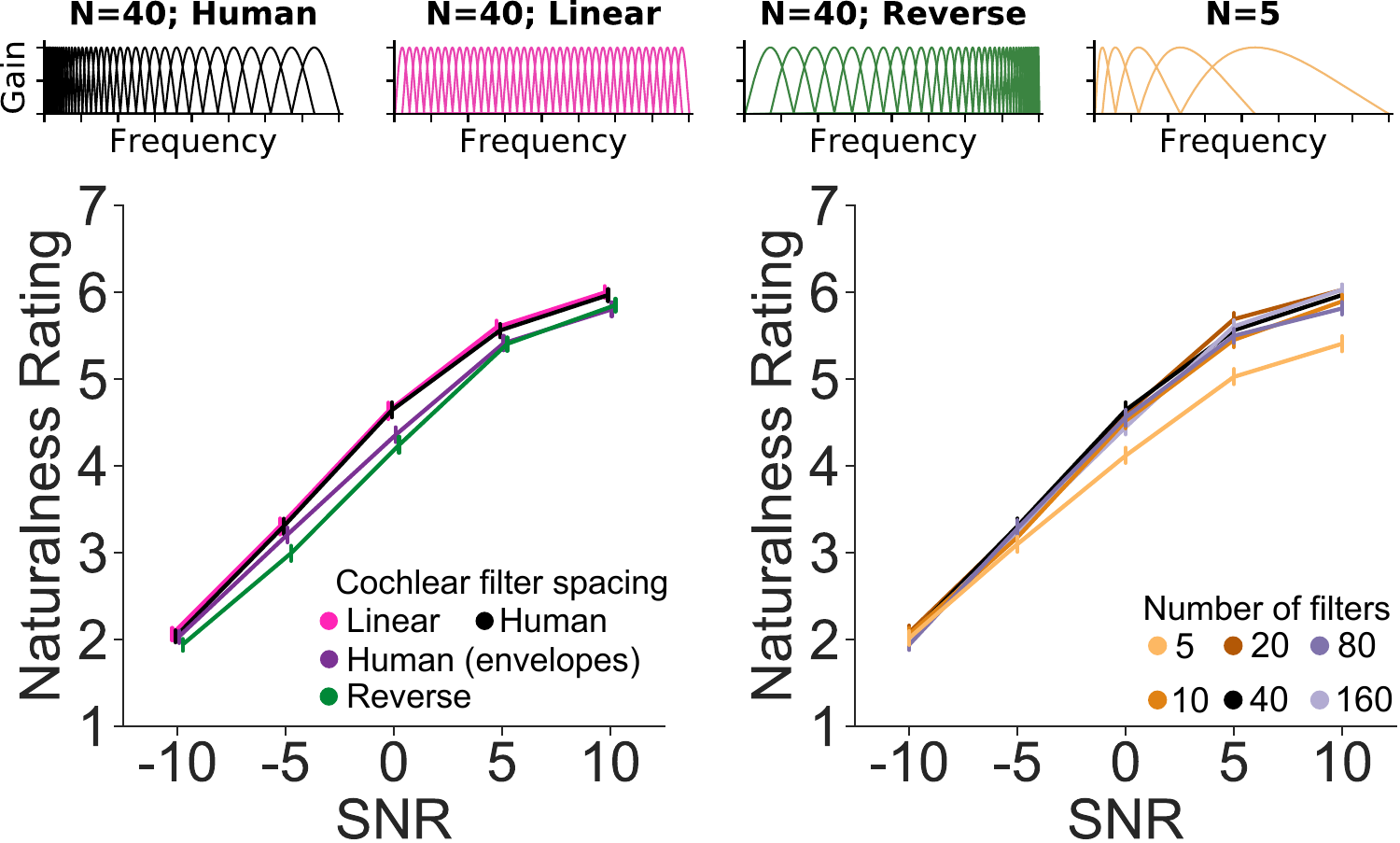}
  \caption{Rated naturalness vs. SNR for speech processed by Wave-U-Nets trained on cochlear model losses with different filter banks (select examples depicted above). Error bars plot SEM (across 105 participants).}
  \label{fig:naturalness_2}
\end{figure}

 The benefit of the cochlear loss depended to some extent on the filter characteristics (Table \ref{tab:expt2}; Figure \ref{fig:naturalness_2}, left). Worse performance was obtained with a `reversed' filter bank, with wide filters at low frequencies and narrower filters at high frequencies, opposite to that of the ear. Using the envelope of the filter outputs also produced worse performance (counter to the hypothesis that phase invariance might be critical). However, filters that were linearly spaced along the frequency axis worked about as well as those modeled on the ear.

 Worse performance was also obtained using only five filters (scaled to cover the frequency spectrum), but good results were obtained provided at least 10 filters were used (Figure \ref{fig:naturalness_2}, right). This result suggests that splitting the audio up into multiple frequency channels is sufficient to replicate the benefit of deep features provided there are enough channels with reasonably sensible frequency tuning.

\subsection{Objective Metrics}
The models trained on deep recognition features also performed better than the baselines according to PESQ and STOI. Notably, this advantage was not evident when measured with SDR. The filter bank-trained models showed the opposite trend -- better performance as measured by SDR, and worse via PESQ and STOI (Table \ref{tab:expt2}). These differences suggest that the filter bank and deep feature losses are not fully interchangeable despite having similar effects on overall naturalness. The results also underscore the limitations of objective metrics for capturing human perception of altered speech.

\section{Discussion}
Prior work has proposed denoising based on deep feature losses \cite{Germain2019,Pilarczyk2019,su2020hifi,hsieh2020improving,kataria2020perceptual}, but has not evaluated it relative to methods using simpler waveform- or subband-based losses. 
We found that deep recognition features could be used to train denoising systems that outperform waveform-based methods, but that their benefit could be matched using a standard one-layer auditory filter bank. The results thus provide no evidence that deep features provide a unique benefit for denoising.


Although deep neural networks yield the best current models of biological sensory systems \cite{yamins2016using,kell2019}, our results indicate that these similarities are not yet sufficient to produce audio enhancement algorithms above and beyond what can be obtained from simple filter bank models. However, it is possible that building better models of human perceptual systems will also yield feature losses \cite{Ananthabhotla2019codecloss,Manocha2020JND} that would better transfer their perceptual benefits to humans, and produce benefits relative to simpler approaches. It also remains possible that the audio quality is limited more by the audio transform than the feature loss. More expressive transforms, or transforms with stronger generative constraints, might yield a clearer benefit of deep features.

The benefits of deep feature and cochlear model losses relative to waveform-based losses were clear from the ratings of human listeners, but were less evident in the objective metrics we tested (PESQ, STOI, SDR). This result indicates that optimizing for auditory model-based losses may provide perceptual benefits that conventional objective metrics are poorly suited to measuring, and suggests to the potential value of auditory model features as new objective metrics.

In sum, we found that audio transforms trained to modify noisy speech so as to reconstruct deep feature representations of clean speech produce better denoising performance than transforms trained to reconstruct clean speech waveforms, as measured by the ratings of human listeners. However, a similar benefit was obtained using one-layer auditory filter banks, suggesting the importance of multi-channel, overcomplete representations rather than learned features per se.

\section{Acknowledgements}
The authors thank Ray Gonzalez for developing the training dataset, John Cohn and the Oak Ridge National Laboratory for use of Summit, and the MIT-IBM Watson AI Lab for funding.

\bibliographystyle{IEEEtran}
\bibliography{main_bib}

\end{document}